\documentclass[twoside]{book}
\usepackage{graphicx}

 \makeatletter
\ifx\UNDEF\mail\def\mail{ }\else\fi
\ifx\UNDEF\prange\def\prange{0 0}\else\fi

\gdef\@empty{}
\def\Mail#1 #2 {\gdef\thecontact{#1}\gdef\theaddr{#2}}
\def\Range#1 #2 {\gdef\thefirstpage{#1}\gdef\thelastpage{#2}}
{\let\'\mail \expandafter\Mail\' }  
{\let\'\prange \expandafter\Range\' }   
 \gdef\@shtitle{\relax}
 \long\def\shtitle#1{\gdef\@shtitle{#1}}
 \long\def\author#1{\gdef\@author{#1}}
 \def\affil#1{\par\noindent{\rm#1\par}}
 \gdef\@abstract{}
 \long\def\abstract#1{\gdef\@abstract{#1}}

 \def\maketitle{\thispagestyle{empty}\chapter{\@title}}
 \renewcommand\chapter{\if@openright\cleardoublepage\else\clearpage\fi
                    \thispagestyle{empty}%
                    \global\@topnum\z@
                    \@afterindentfalse
                    \secdef\@chapter\@schapter}
 \def\@makechapterhead#1{%
  \vspace*{50\p@}%
  {\parindent \z@ \raggedleft \normalfont
    \ifnum \c@secnumdepth >\m@ne
      \if@mainmatter
        \par\nobreak
        \vskip 20\p@
      \fi
    \fi
    \interlinepenalty\@M
    \Huge \bfseries #1\par\nobreak
    \vskip.25in
    \large\bfseries\@author\par\nobreak
    \vskip 40\p@}
    \ifx\@abstract\@empty\else{\small\@abstract\par\vskip20\p@}\fi
  }

\DeclareRobustCommand\em
        {\@nomath\em \ifdim \fontdimen\@ne\font >\z@
                       \upshape \else \slshape \fi}

\def\@begintheorem#1#2{\sl \trivlist \item[\hskip \labelsep{\bf #1\ #2}]}
\def\@opargbegintheorem#1#2#3{\sl \trivlist
     \item[\hskip \labelsep{\bf #1\ #2\ (#3)}]}

 \newcommand{\sectlabel}[1]{\label{sect:#1}}

  \def\@arabic#1{\number #1} 

\long\def\@makecaption#1#2{
    \vskip\abovecaptionskip
    \sbox\@tempboxa{{\small {\bf #1}: #2}}%
    \ifdim\wd\@tempboxa>\hsize
        {\small {\bf #1}: #2\par}
    \else
       \global\@minipagefalse
       \hbox to\hsize{\hfil\box\@tempboxa\hfil}
    \fi
    \vskip \belowcaptionskip}

\def\figstrut#1{\hbox to\linewidth{\vrule height#1\hfill}}

\renewenvironment{thebibliography}[1]
     {\section*{\bibname
        \@mkboth{\MakeUppercase\bibname}{\MakeUppercase\bibname}}%
      \list{\@biblabel{\@arabic\c@enumiv}}%
           {\settowidth\labelwidth{\@biblabel{#1}}%
            \leftmargin\labelwidth
            \advance\leftmargin\labelsep
            \@openbib@code
            \usecounter{enumiv}%
            \let\p@enumiv\@empty
            \renewcommand\theenumiv{\@arabic\c@enumiv}}%
      \sloppy
      \clubpenalty4000
      \@clubpenalty \clubpenalty
      \widowpenalty4000%
      \sfcode`\.\@m}
     {\def\@noitemerr
       {\@latex@warning{Empty `thebibliography' environment}}%
      \endlist}
\makeatother

 \title{Dynamical Emergence of Complex Structures in Field Theories}
 \author{Joel Thorarinson \& Marcelo Gleiser \affil{Department of Physics and Astronomy \\ Dartmouth College,
\\ Hanover, NH 03755, USA\\thorvaldur@dartmouth.edu \\ gleiser@dartmouth.edu}}
 
 \abstract{
 Nonlinear field theories can be used to study both standard physics
questions, or to study questions such as the emergence of order and
complexity. These theories are generally derived from the symmetries
of a given problem and the interactions that respect those
symmetries. Formally one can then quantize the system to find the
masses of the fundamental excitations, but this procedure generally
destroys much information about solutions of the field equations
with large non-perturbative amplitudes. To get information about the
properties of the solutions to these field theories without
perturbative approximations we use real time lattice simulations
where complex spatiotemporal structures emerge dynamically from
excess free energy and a thermal background. We present results in 2
and 3 dimensions of new interacting quasi-particle formations from a quench. These objects show an emergent level of complexity which we attempt to categorize and define in a manner that should be useful to many different applications.
 }

\def\naive{na\"{\i}ve }

\def\L       {{\mathcal L}}
\def\D       {{\mathcal D}}

\def\pd      { \partial}
\def\V       {{\mathcal V}}

\def\L       {{\mathcal L}}

\def\pd { \partial}
\def\V       {{\mathcal V}}
\def\bra     {{\langle}}
\def\ket     {{\rangle}}
\def\a{\alpha}
\def\b{\beta}

\def\be{\begin{equation}}
\def\ee{\end{equation}}
\newcommand{\eea}{\end{eqnarray}}
\newcommand{\bea}{\begin{eqnarray}}

\def\d{\delta}
\def\l{\lambda}
\def\d{\delta}

\def\l{\lambda}

\def \om { \omega}

\def \p { \phi}

\def\pdbk{\bra \phi^\dag \ket}
\def\pbk{\bra \phi \ket}

\def\V2{\mu^2 \pdbk \pbk + \lambda (\pdbk \pbk )^2 }

\begin{document}          
\maketitle
                           
\section{Introduction}\sectlabel{intro}	

Although our Universe can be very accurately described by simple microscopic equations, the macroscopic level phenomena appear significantly different. From this, we can infer that distinct emergent levels of complexity is a fundamental property of Nature \cite{anderson}. In this paper, we will discuss the basic characteristics of some systems that show emergent levels and how these levels can emerge dynamically. We will also discuss the relation between the definition of these levels and the definition of complexity. These relations can then give us insight into the dynamics of emergent levels of complexity in a general set of similar theories and physical systems. Since we are looking for a description of levels of complexity that is closely related to our Universe, we start with equations similar to those found in the Standard Model of particle physics as opposed to models such as cellular automatons. Based on definitions of emergence and complexity, we will show numerically that distinct levels of complexity can emerge dynamically from a single low level (fundamental) description of a system which has emergent higher level descriptions (effective).  

A system in which there is a fundamental, and multiple effective levels of mathematical description, represents a good theoretical laboratory into the relationships between these levels, if any. Usually we take complex systems and break them down into simpler constituent parts in order to understand the pieces, and then the whole. We have found a few exemplary systems in which we can start from a known basic theory and build effective descriptions for the emergent levels. The evidence for the existence of emergent levels was obtained numerically. In practice, this means that we can start bottom-up, from a fundamental description to a complex effective one. This allows us to investigate the possible relationship between a systems different realizations and their mutual correspondence.

\paragraph{Observational definitions of emergent complexity: }

Our working definition of complexity will be that the system should have different levels of description. These levels can be seen by identifying three fundamental characteristics. The first we call {\it separability}, which means that when looking at a system one should be able to define separately the characteristics of each level without reference to the others in some region of configuration space. 

The second is that there should exist a region in configuration space between levels where the phenomena are intrinsically linked and one cannot make a distinct level characterization. This we call {\it level mixing}. We conjecture that this property of {\it level mixing} could be the principal source for what S. Wolfram has called ``Class 4 behavior," which cannot be completely characterized as random or repetitive;\cite{wolfram:2002} but shares both properties. This property is essential for the characterization of lower level ``living" or ``intelligent" systems. It also has the power of explaining the dynamics of genetic evolution through the interaction of two levels, one that stabilizes information and one that perturbs the informational structures to find more efficient formations. This interaction allows the dynamical evolution of genetics without resorting to purely probabilistic interpretations and gives a solid although complex reason for the persistence of localized genetic structures such as in living systems. 

The third characterization of complexity is a phase transition in the mathematical information necessary to solve or analytically handle the system. The potential for {\it algorithmic complexity} in a system is an essential and powerful element for understanding both the potentialities of a system and our own potential for extracting physical information from it using known techniques. We will discuss these definitions further with examples in section \ref{emcft}.

The idea of emergence is also essential to any definition of complexity which hopes to reconcile the non-disipative flow of information from micro to macroscopic scales. Emergence can be defined operationally as the property of a system with a given energy to flow towards a nontrivial region in phase space. Thus, emergent configurations will be those that are attractors in phase space and which do not satisfy equipartition. For example, a configuration which remains dominantly in one region of phase space where for its given energy there exist many other equally probably states could then be considered emergent. Observationally, this means that we expect to see new localized effective quasi-particles which are macroscopic excitations of many fundamental quanta. The interactions that generate these quasi-particles also generate new length scales and effective interactions which characterize a multi-scale system. Effective multi-scale dynamics is an element that has also been used to characterize emergent complex behavior \cite{baryam}.

Emergent behavior as we have defined it generally comes from out of equilibrium dynamics. Care is needed in defining equilibrium, equipartition is when each independent quadratic term in the energy functional carries the same averaged energy. This is basically a global phase space equilibrium that does not support nonperturbative complexity. The spatially extended configurations we are interested in for this paper are local phase space equilibria. This means that they can hold energy in a dynamic equilibrium in some region of phase space. This also implies that there is an effective barrier in configuration space, preventing the structure from dissipating toward equipartition. Understanding a systems potential for generating finite, spatially-extended structures is thus important in gauging the possibility that it has emergent properties.

Characterizing the local equilibrium and the effective barrier in configuration space is very complicated analytically due to the nonlinear nature of the problem. Most often we propose semi-analytic justifications for various phenomena. This is intrinsically linked to the systems potential for {\it algorithmic complexity}. Having analytically unsolvable systems is actually not a problem, it is Nature's way of telling us to generate alternative approaches to study nonlinear dynamics.  Currently our most useful tool is large-scale numerical simulations.

\section{Emergent Complexity in Field Theories } \label{emcft}

We will work directly with two relativistic field models in the next sections, the Abelian Higgs model (AH) and the real scalar field model with a double well potential (RDW). The AH has a Lagrangian density of:
\be
\label{Lagden}
 \L=\D_\mu \phi^\dag \D^\mu \phi - \frac{1}{4}F^{\mu \nu}F_{\mu
\nu}-V(\phi^\dag \phi), 
\ee where $\D_\mu=\pd_\mu-igA_\mu$, $F_{\mu \nu}=\pd_\mu A_\nu - \pd_\nu A_\mu$ and $V=\frac{\l}{4}(\phi^\dag\phi-\eta^2)^2$.  The scalar and vector masses are, respectively, $m_s=\sqrt{\lambda}\eta$ and $m_v=\sqrt{2}g\eta$. Their ratio defines the parameter $\beta=(m_s/m_v)^2=\lambda/2g^2$ which is one measure of the lowest level multi-scale potential of the system. The RDW model can be obtained from the AH model by setting $A_\mu=0$, $g\equiv 0$, $V(\p)=\frac{\p^2}{2}-\a\frac{\p^3}{3}+\frac{\p^4}{8}$, where $\p$ is a real valued field. For a greater understanding of the scalings, numerical implementation, and equations of motion see ref. \cite{Gleiser:2007te} for the AH model and ref. \cite{Gleiser:2007ts} for the RDW model.

We have good reason to believe that these models have the ability to generate non-trivial spatiotemporal patterns or structures which are themselves coherent packages of many fundamental quanta. In both the AH and RDW models we see localized large amplitude coherent oscillatory structures which are superimposed on top of the high $k$ dynamics of the field. The localized configurations have been identified as oscillons \cite{D-dim} and much work has been done to attempt to understand them due to their ubiquity in nonlinear field theories such as the $SU(2)XU(1)$ \cite{Graham:2006vy} electroweak model and also the $U(1)$ AH model \cite{Gleiser:2007te}. 

These oscillatory structures probe the nonlinear parts of the potential. It has been shown that for RDW models that they must have an inflection point on the potential in order for oscillons to exist\cite{D-dim}. This condition is very interesting as it guarantees that the potential has a negative effective mass at some point. If one decomposes the field into a basis $\phi(t,x)\propto e^{i\om t} f(x)$, then, because the frequency of oscillation of the amplitudes will be related to the mass by $\om\sim m_{eff}$, there must be instabilities in the fields amplitudes.  

These instabilities have three important consequences when considering how to characterize the complexity of the system. The first is in relation to the {\it algorithmic complexity} of solving the system. Generally, what we mean by solving a system is quantization, which amounts to (simplistically speaking) finding the eigenmodes. In this case, the basis we usually use (the modes from the Fourier transform) is unstable due to the non-analytic mass. This means that any attempt to decompose the lower momentum spectrum in this basis will necessarily fail as the instability in the amplitudes will promote the continual mixing of modes less than $k \sim m_{eff}$. In practice, this means that we are unable to find the proper eigenfunctions for the system and thus must resort to numerical studies. What we find is a very rich structures, characterized by the nonlinear dynamics of low momentum, large amplitude modes.

The second consequence is that the instability drives the system to a new emergent metastable equilibrium represented by a condensate of correlation amplitudes at lower momentum and spatially in structures identified as oscillons. This gives us a very important criterion to evaluate emergent systems: they should have \naive  instabilities and apparent non-analyticities. Amazingly, this is where linear intuition is useless and the magic of nonlinear emergence is manifest: instabilities in the presence of sufficient free energy generate their own effective equilibrium structures that  look nothing like the propagating waves of the perturbative analysis. The third consequence of generating an effectively multi-scale system we address with an example when we talk directly about the AH model.


\paragraph{Configurational complexity:} For another example of {\it algorithmic complexity} consider the information necessary to characterize a system or configuration. If a system has no spatiotemporal variation then the characterization at one point will suffice for all points and very little information is needed. The other extreme is that there is variation at every point in space as in a thermal configuration. This can also be characterized (probabilistically) by a single number, such as the temperature, which labels a distribution. 

In between these two extremes is the region where there are emergent spatiotemporal structures. They need information to be described individually and collectively. Somewhere between the two extremes we could say that the system has maximum complexity. Again, informational characterizations of complexity such as Shannon's entropy \cite{Shannon:1948} can be a bit deceiving because although the thermal state would take the maximum number of bits to exactly characterize, much of that information is useless and can be handled probabilistically with simpler distributions labeled by the  temperature. So, the maximum complexity of a physical system lies not in the maximum bits to characterize it but in the maximum algorithmic steps necessary to describe the system's spatiotemporal structures after removing those quantities that can be labeled by global values and distributions.

\begin{figure}
\begin{minipage}{0.5 \linewidth}
\centering
\includegraphics[scale=0.79,angle=0]{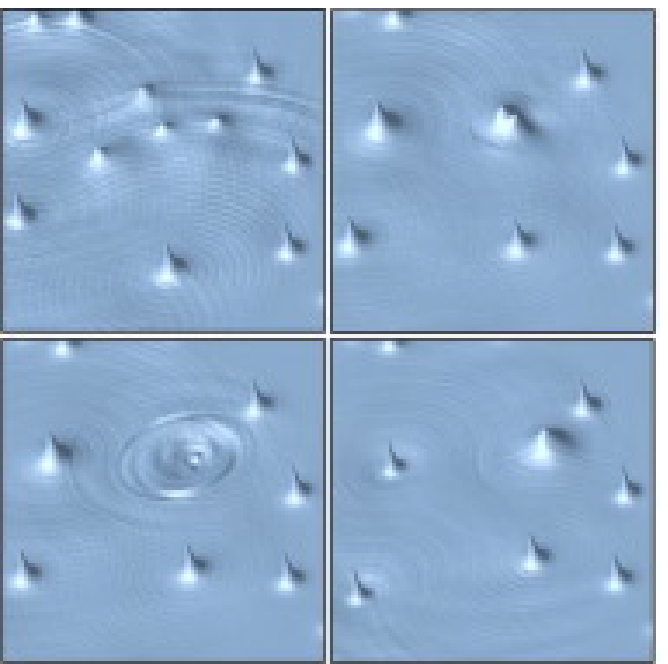}
\end{minipage}%
\begin{minipage}{0.5 \linewidth}
\centering
\includegraphics[scale=.44,angle=0]{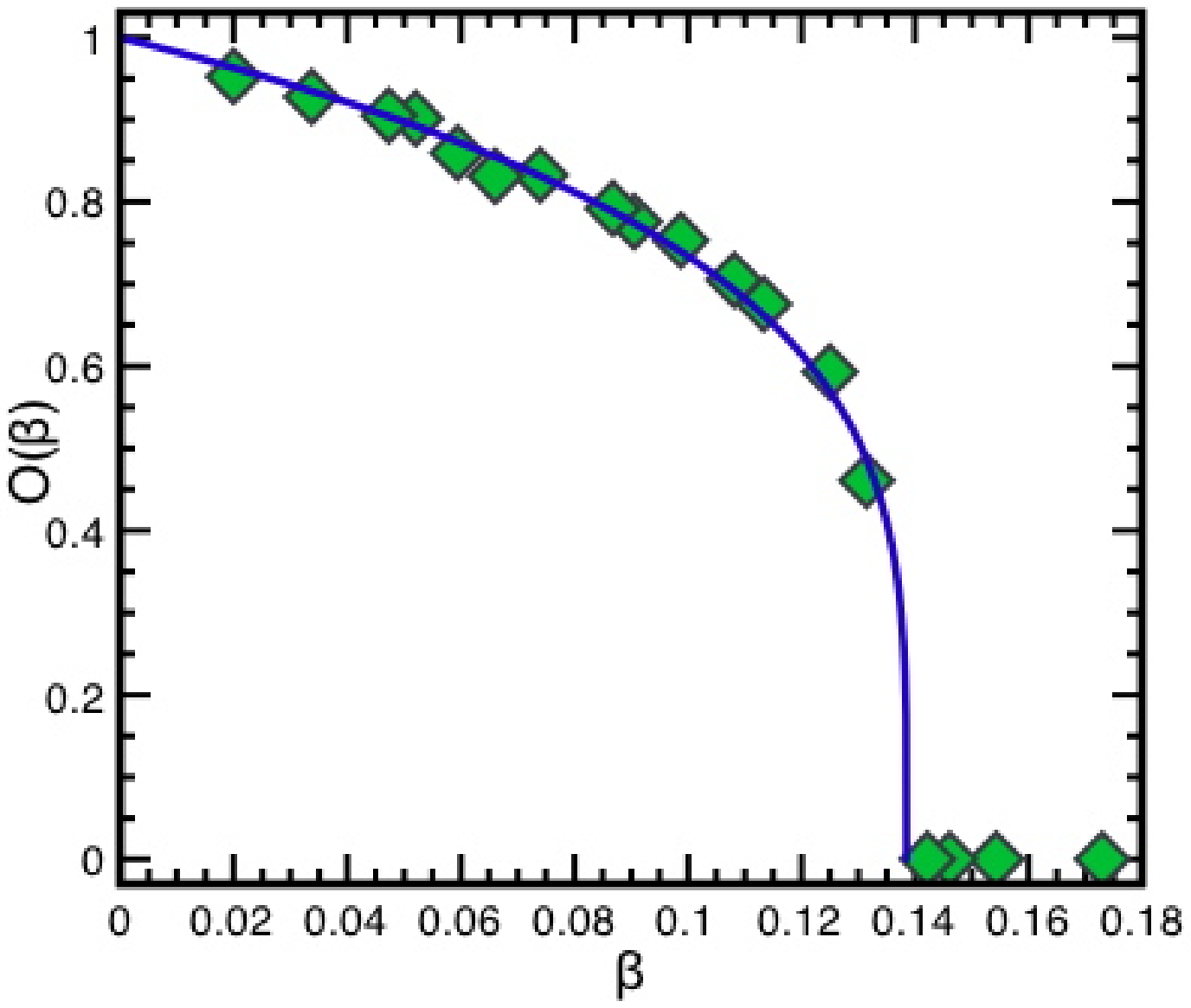}
\end{minipage}%
 \caption{\label{fig:vorcon}  Left: snapshots of vortices and oscillons at four times for $\beta=0.04$. The smaller oscillons condense and radiate to form the larger metastable oscillon in the bottom right corner. Right:  fraction of oscillon to vortex-antivortex (vav) energy as a function of mass ratios $\b$ This shows that the emergence of stable oscillons from vav annihilation can be treated as a critical parameter and that their existence is preferred in the region of parameter space which has two strongly different mass (length) scales.} 
\end{figure}

\paragraph{Complexity in the AH model:}The AH model gives us a direct example that exhibits both analytic and configurational {\it algorithmic complexity}. One can also easily identify {\it separability} and {\it level mixing}, as well as both basic and emergent multi-scale behavior. 

The basic multi-scale parameter is the pertubative level definition of the effective mass ratio $\beta$. The emergent multi-scale parameter is the effective radius of the quasi-particle oscillons that form from the annihilation of two topological particles (vortices). We have seen on the lattice that near the critical mass ratio $\beta_c \sim 0.136$ that the radius of the oscillon goes as $R_{osc}\sim (\beta-\beta_c)^{-1}$. This $R_{osc}$ then generates another emergent interaction scale between quasi-particles for that region of parameter and phase space.

{\it Separability} and {\it level mixing} can be easily identifiable as the emergent quasi-particle structures such as vortices and oscillons (see fig.\ref{fig:vorcon}) at one level and the thermal linear modes at the more fundamental level. The interaction between these levels generates very complex dynamics and emergent behaviors such as oscillon coalescence, in which multiple oscillons interact to form a larger composite quasi-particle. This is one example where microscopic many-body interactions of fundamental modes interact to create two forms of quasi-particle, each of which with the ability to interact with each other to form larger structures and thus to generate the third possible level of complexity. The {\it algorithmic complexity} is manifest for reasons we discussed above. Systems such as the AH model which exhibit three levels of complexity from one fundamental set of equations certainly deserve a more detailed investigation. One would hope to find the dynamical principles that direct the flow of complexity between levels.

\begin{figure*}
\centering
\begin{minipage}{0.5 \linewidth}
\includegraphics[scale=0.44,angle=0]{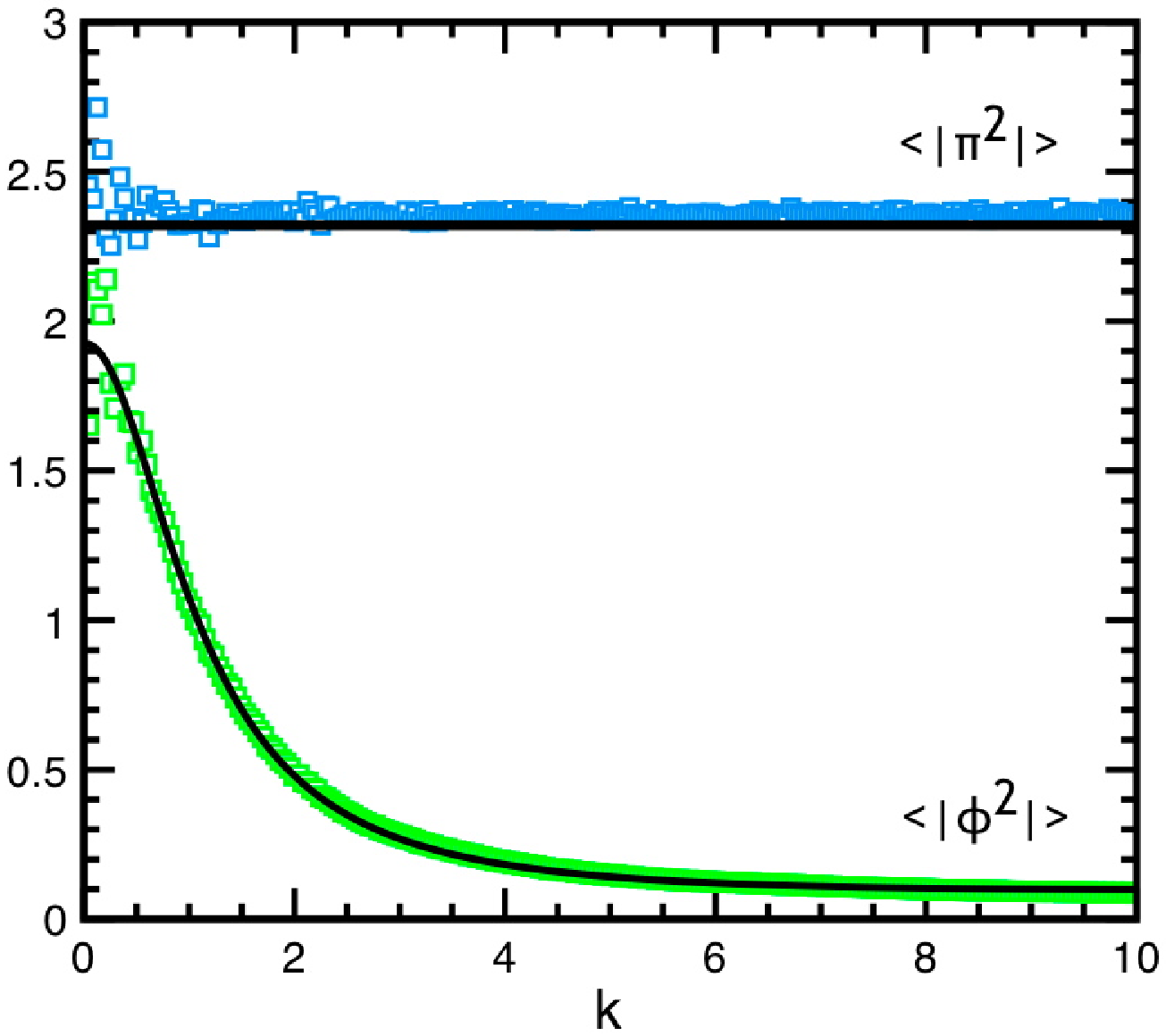}
\end{minipage}%
\centering
\begin{minipage}{0.5\linewidth}
 \includegraphics[scale=0.44,angle=0]{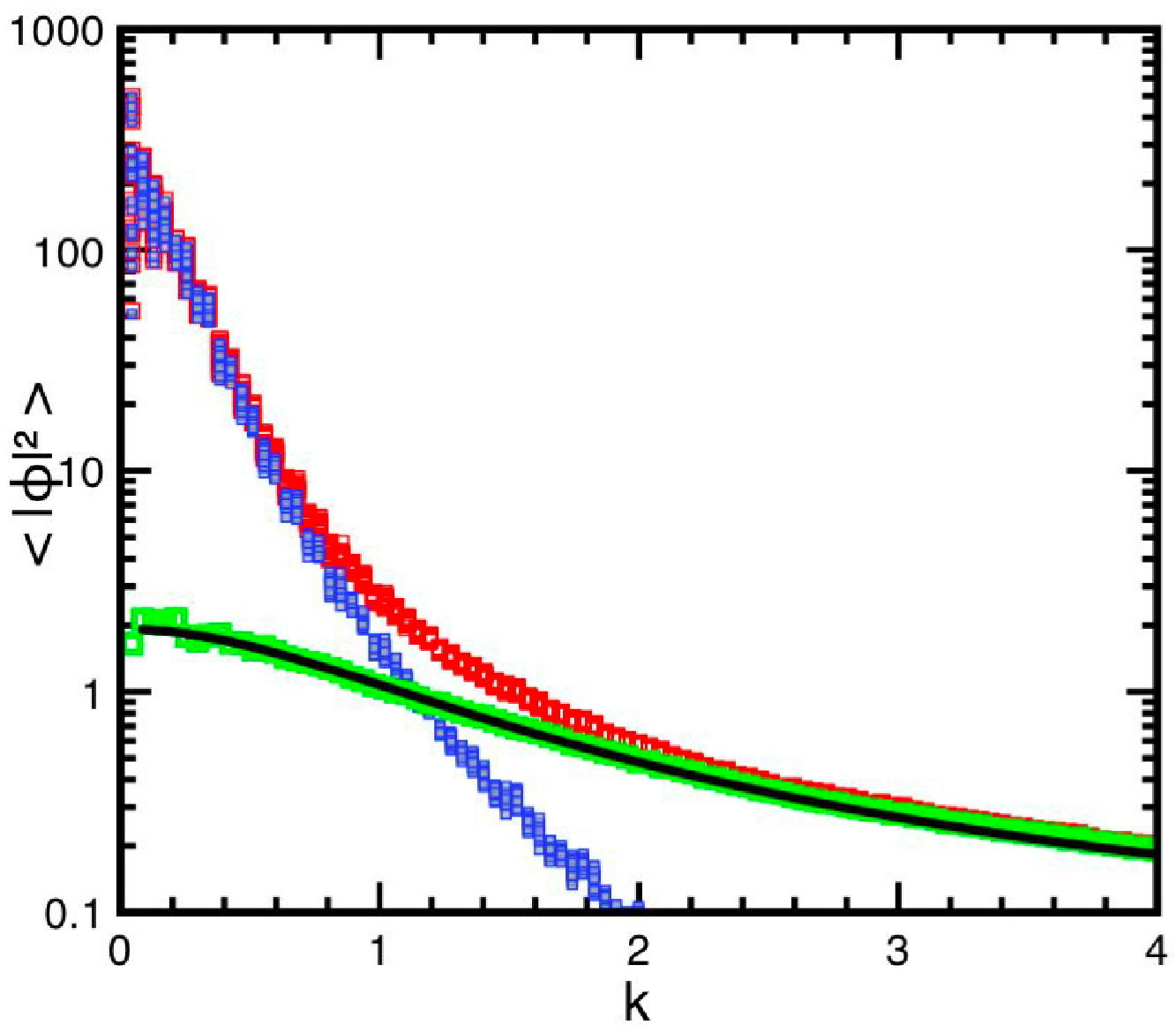}
\end{minipage}
 \caption{\label{corr_osc} The radially-averaged 2-point field and momentum correlation functions. On the left, we show the results for the thermal initial conditions. The black continuous lines correspond to $\langle |\bar{\pi}(k)|^2\rangle =T (\d x)^{-3}$ (top) and to  $\langle |\bar{\phi}(k)|^2\rangle = {T \d x^{-3}}/\{k^2+m_{\rm H}^2\}$ (bottom).The blue and green squares are from the simulation. On the right, we plot the logarithm of the two-point correlation function for the field. The black and green correspond to the thermal initial state, as in the plot on the left. The red squares are the data plotted at every half second for the time interval $360<t<400$. The blue squares are also the data, after applying a thermal filter.  One clearly distinguishes two populations of modes: the oscillon-related, low $k$ modes ($k\leq 1.5$), which sharply depart from the thermal spectrum, and the purely thermal modes, which remain in thermal equilibrium throughout the simulation ($k\sim 1.5$). These figures were generated using the data from the same run as in \ fig.\ref{symosc}.}
 \hspace{1.5 cm}
 \centering
\begin{minipage}{1.0\linewidth}
 \centering
 \includegraphics[scale=.48,angle=0]{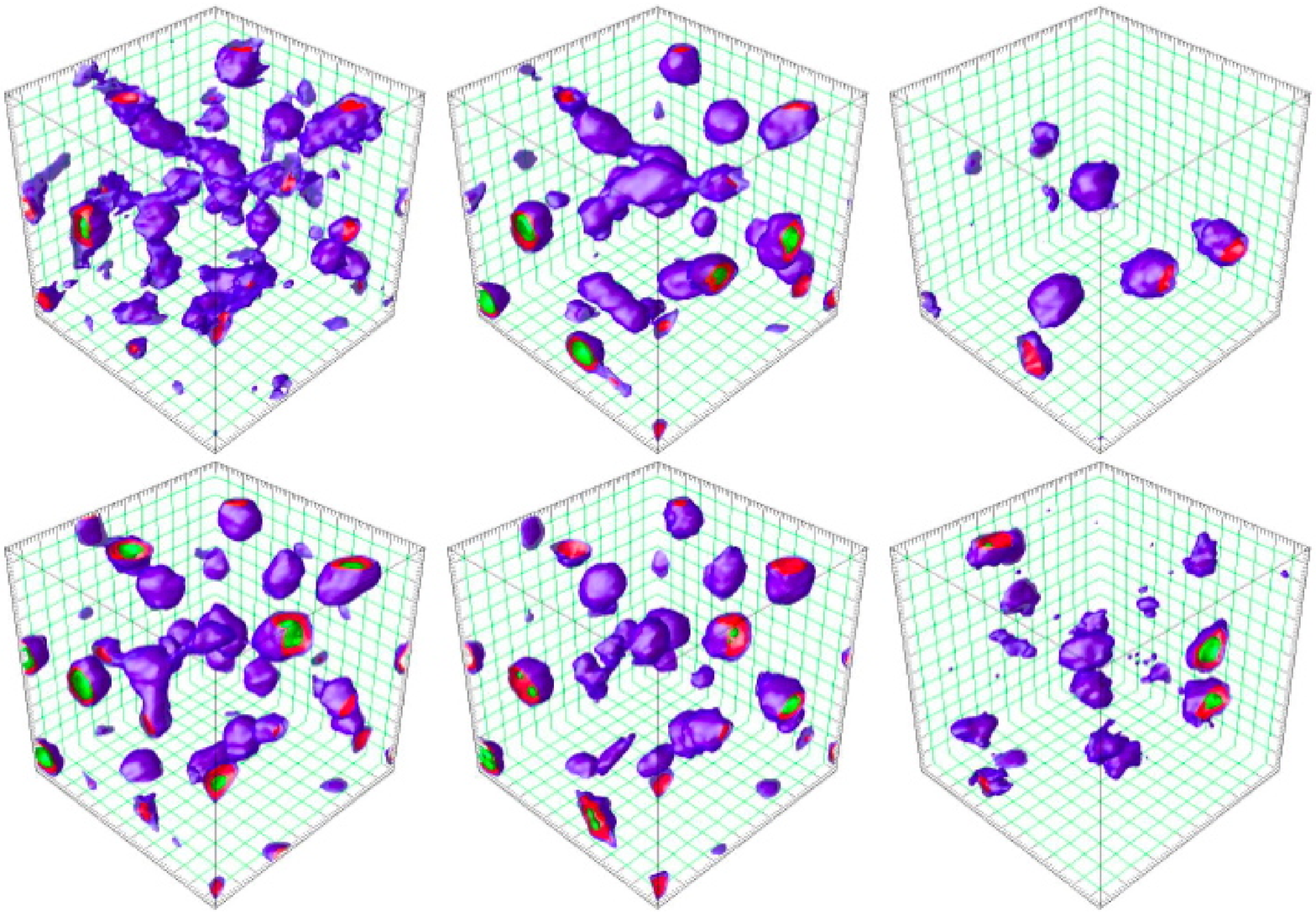}
\end{minipage}
  \caption{\label{symosc} Snapshots of a $3d$ scalar field simulation after a quench from a single to a symmetric double well potential ($\a=0\rightarrow1.5$). The oscillatory bubble-like configurations which emerge in synchrony are identified as oscillons. The lattice parameters where $dx=0.5$, $T=0.29$ and $V=64^3$. Plotted is the $\phi=\{0.75,1.0,1.5\}$ isosurfaces in blue, red and green respectively.  The plots are instants of time taken at time $t=\{45.5s,51.5s,54.5s,60.5s,62.0s,194.5s\}$ respectively increasing left to right and top to bottom. } 
 \end{figure*}

\paragraph{Emergent populations in the RDW model:}Using the RDW model, we will now give an example of a system with two levels of mode distributions emerging from the same underlying theory.  To generate these populations we first heat the real scalar field in a quadratic potential ($\a = 0$) about $\phi=0$ to some temperature $T$ by using Langevin dynamics (see \cite{Gleiser:2007ts,Batrouni:1985jn}). We know that the analytic momentum space correlations of the field $\phi(k)$ are labeled simply by $T$ and the effective mass of the field $m_{eff}(T)$. After preparing this simple state as an initial condition we then quench the field to a double well potential ($\a \rightarrow 1.5$). This generates fluctuations in the zero mode or (volume-averaged) field $\bra \p(t)\ket=\frac{1}{V} \int d^{3} x \p$, which effectively dumps free energy into the lower momentum modes. Eventually this free-energy will distribute itself evenly about the high $k$ modes as the system equilibrates to a new temperature.

Amazingly, there is much that happens on the way. The global expectation value of the field does settle down into its effective minimum after oscillating for a bit. However, the correlation values are drastically different from what one would expect for a thermal distribution. In fig.\ref{corr_osc} we show the radially and ensemble-averaged distributions for the field ($\bra |\p(k)|^2\ket$) and the momentum ($\bra |\pi(k)|^2\ket$) on the left are the initial thermal values. On the right we show the fields correlations for the initial values (green) again and for the the asymptotic ($360 < t < 400$) times (red) where the  $\bra \p(t)\ket$ appears to be in equilibrium. We have also plotted in blue $\bra |\p(k)|^2\ket$ with the initial thermal spectrum removed. Note that this new nonlinear spectrum (fig.\ref{corr_osc} blue) is amplified by around three orders of magnitude over the thermal spectrum at low momentum.  What this shows is that there are two simultaneous weakly-interacting metastable equilibrium quasi-particle species within the same theory.  We have a theory for the linear (thermal) modes but the we are just beginning to explore the nonlinear modes. The apparent decoupling between these modes and the high-$k$ spectrum adds strength to our claim that this represents two populations, one thermal and (high-$k$), and one emergent population obeying different statistics and quite stable. This example fulfills our criterion of complexity in that there is apparent {\it separability}. Identifying {\it level mixing} in this theory is a more complicated question that we do not have space to address here.

In fig.\ref{symosc} we plot the actual field isosurfaces after the thermal spectrum has been removed (see caption for details). What we see are localized large amplitude coherent oscillatory structures which are superimposed on top of the high-$k$ dynamics of the field. These meta stable structures represent the new long-lived non-propagating modes which have emerged. Moreover, they are apparently only weakly coupled to the thermal spectrum of the field and one can isolate each level clearly in Fourier space. So for the RDW we have shown that the complexity levels have both {\it separability} and {\it algorithmic complexity}.  For a review on how drastically these emergent structures can modify the dynamics of the system see ref.\cite{Gleiser:2007ts}.

\section{Conclusion } \label{conclusion}

We have presented two examples where emergent complexity can be studied from a fundamental theory with easily identifiable levels and spatiotemporal coherent phenomena. Also, we have proposed definitions that are easily applicable to the question of emergence in nonlinear physical systems. We have briefly argued that the principles of {\it separability}, {\it level mixing}, and {\it algorithmic complexity} can be useful in gaining perspective on a wide range of physical and, possibly biological phenomena.

\end{document}